\documentclass[aps, prd, showpacs, nofootinbib, twocolumn]{revtex4}
\usepackage{amssymb,amsmath,microtype}
\usepackage{graphicx}
\usepackage[normalem]{ulem}

\usepackage{color}

\def\la{\left(}
\def\rf{\right)}

\begin{document}
\title{ Probing the gauge symmetry breaking of the early universe in 3-3-1 models and beyond by gravitational waves}
\author{Fa Peng Huang}
\author{Xinmin Zhang}

\affiliation{Theoretical Physics Division, Institute of High Energy Physics, Chinese Academy of Sciences, P.O.Box 918-4, Beijing 100049, P.R.China \\
School of Physics Sciences, University of Chinese Academy of Sciences, Beijing 100039, China}

\begin{abstract}

Taking the 3-3-1 models (with $SU(3)_c \otimes
SU(3)_L \otimes U(1)_Y$ gauge group) as examples,  we study that a class of new physics models with extended gauge group could undergo 
one or several first-order phase transitions  associated with the spontaneously symmetry breaking processes during the evolution of the universe,
which can produce detectable phase transition  gravitational wave (GW) signals at future GW experiments, such as LISA, BBO, DECIGO, SKA and aLIGO.
These GW signals can provide new sources of GWs with different peak frequencies, and can be used to probe the evolution history of the universe.
\end{abstract}


\maketitle

\section{Introduction}
The observation of gravitational waves (GWs) by Advanced Laser Interferometer Gravitational Wave Observatory (aLIGO)~\cite{Abbott:2016blz} has initiated a new era of exploring the cosmology, the nature of gravity as well as the fundamental particle physics by the GW detectors~\cite{Schwaller:2015tja,Dorsch:2014qpa,Huang:2016odd,Dev:2016feu,Jaeckel:2016jlh,Yu:2016tar,Addazi:2016fbj,Huang:2016cjm}.
Especially, due to the limitation of the colliders' energy, GW detectors can be used as new or complementary techniques to probe the existence of the new physics (NP) by detecting the symmetry breaking patterns or phase transition history for large classes of NP models with an extended gauge group, which are motivated by the mysterious experimental results in our understanding of
particle cosmology (such as the dark matter problem or the puzzling observed baryon asymmetry
of the universe), and the absence of NP signals at current collider experiments.
The increasingly attractive NP models with an extended gauge group have many new particles
without leaving obvious observable imprints at current particle colliders.
However, the GW experiments may provide a possible approach to test their existence.
For example, to explain the baryon asymmetry of the universe via electroweak (EW)  baryogenesis, a strong
first-order phase transition (FOPT) is needed to realize the departure from thermal equilibrium  by extensions of the standard model (SM)~\cite{Huang:2015bta,Huang:2015izx,Vaskonen:2016yiu}.
And during the FOPT, detectable GWs will be produced through three mechanisms:
collisions of expanding bubbles, sounds waves, and magnetohydrodynamic
turbulence of bubbles in the hot plasma~\cite{Witten:1984rs, Hogan:1984hx, Turner:1990rc,Kamionkowski:1993fg,Hindmarsh:2013xza,Hindmarsh:2015qta,Kosowsky:2001xp,Caprini:2009yp}.
Phase transitions in particle physics and cosmology are usually associated with the symmetry breaking, i.e.
where the universe transits from a symmetric phase to a symmetry broken phase when
the temperature drops below the corresponding critical temperature.

For the first time, we have a realistic chance to explore NP with
gauge symmetry breaking processes through phase transition GW signals
after the discovery of the GWs by aLIGO, which is particularly exciting.
In this paper, we study the possibility to probe the gauge symmetry breaking patterns
and the phase transition history of the early universe by the phase transition GW signals.
In particular, we focalize our analysis to GW detection of the NP models with an extended non-Abelian gauge group,
where the symmetry breaking at each energy scale
may associate with a FOPT, as shown in Fig.\ref{sb}.
The group $G_{\rm Hidden}$ can spontaneously break into the SM gauge group
via one or several steps and strong FOPT can
take place in each step, which can produce detectable phase transition GWs.
For example, the gauge group $G_{\rm Hidden}$ can be the non-Abelian gauge group $SU(3)_c \otimes
SU(3)_L \otimes U(1)_Y$, which is called 3-3-1 model~\cite{Pisano:1991ee,Frampton:1992wt}.
We show that many versions of the 3-3-1 model can produce at least one strong FOPT at TeV scale in some parameter spaces, which
can produce detectable GW spectrum by the recently proved Laser Interferometer Space Antenna (LISA)~\cite{Seoane:2013qna,Audley:2017drz}, Big Bang Observer (BBO)~\cite{Corbin:2005ny}, Deci-hertz Interferometer Gravitational
wave Observatory (DECIGO)~\cite{Seto:2001qf,Moore:2014lga}, and Ultimate-DECIGO~\cite{Kudoh:2005as}.
In general, there can exist several spontaneous symmetry breaking processes in NP models,
which may also accompany several FOPTs with the evolution of the universe as shown in Fig.\ref{sb}.
If the scale of the FOPT associated with the symmetry breaking is about
$10^7-10^8$ GeV, the phase transition GW spectrum
may be within the sensitivity of future aLIGO.
If the strong FOPT occurs at the QCD phase transition scale in some hidden QCD models~\cite{Schwaller:2015tja}, the
produced GW signals may be tested by pulsar time array (PTA) at the Square Kilometre Array (SKA)~\cite{Smits:2008cf} or the Five-hundred-meter Aperture Spherical Telescope (FAST)~\cite{Hobbs:2014tqa}.

This paper is organized as follows:
In Section II, we schematically discuss the GW detection of the gauge group symmetry breaking and
show how to calculate the phase transition GWs during the FOPT.
In Section III, we will study the phase transition GW spectra in some concrete models with extended gauge group. 
In Section IV, we show our final discussions and conclusions.

\section{First-order Phase transition gravitational wave spectrum}
\begin{figure}
\begin{center}
\includegraphics[scale=0.34]{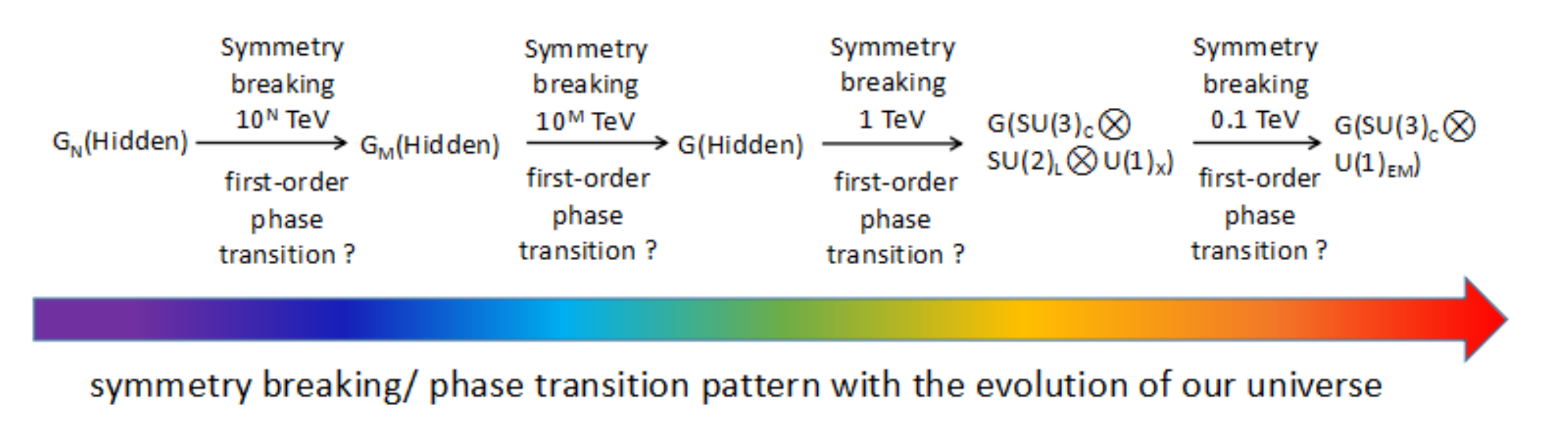}
\caption{Symmetry breaking (phase transition) patterns in the NP models  with extended gauge group  during  the evolution of the early universe, where FOPT may occur.}
\label{sb}
\end{center}
\end{figure}

In a generic classes of NP models, one or several cosmological phase transitions can occur during each step's
symmetry breaking at different energy scale with the evolution of the universe as shown in Fig.~\ref{sb}.
For example, the symmetry breaking pattern may be
$G(\rm Hidden N)\cdots \to G(\rm Hidden 1) \to G(SU(3)_C \otimes SU(3)_L \otimes U(1)_Y) \to G(SU(3)_C \otimes SU(2)_L \otimes U(1)_X)\to G(SU(3)_C \otimes U(1)_{\rm EM})$.
With the evolution of our universe, symmetry breaking will happen at corresponding energy scale,
where the strong FOPT may take place.
Detailed models are given in Section III.

With the evolution of the universe,
the universe transits from a `false' vacuum to a `true' vacuum, and strong FOPT
occurs if there exists a sufficient potential barrier between them.
These processes can produce observable stochastic GW signals, which can be detected in some GW detectors,
such as aLIGO, LISA, BBO, DECIGO, Ultimate-DICIGO, SKA, FAST and so on. Their sensitivity range for some critical
temperatures depends on the energy scale
of the FOPT for different gauge group extended models, as shown in Section III.
To discuss the GW spectra from FOPT, it is necessary to begin with
the one-loop finite temperature effective potential $V_{\mathrm{eff}}(\Phi,T)$:
\begin{equation}
V_{\mathrm{eff}}(\Phi, T) = V_{\rm tree}(\Phi)+V_{\rm cw}(\Phi) +
V_{\rm ther}(\Phi, T) +V_{\rm daisy}(\Phi, T), 
 \label{veff}
\end{equation}
where $\Phi$ represents the order parameter field for the phase transition,
$V_{\rm cw}$ is the one-loop Coleman-Weinberg potential at $T=0$, and $V_{\rm ther} +V_{\rm daisy}$
is the thermal contribution including the daisy resummation~\cite{Quiros:1999jp}.
During each step of symmetry breaking in the NP models  with extended gauge group,
strong FOPT may occur. During a strong FOPT, bubbles are nucleated via quantum tunneling or thermally fluctuating the potential barrier with the nucleation rate per unit volume $\Gamma = \Gamma_0(T) {\mathrm e}^{-S_E(T)}$ and $\Gamma_0(T)\propto T^4$~\cite{Linde:1981zj}, where $S_E(T)\simeq S_3(T)/T$ is Euclidean action~\cite{Coleman:1977py,Callan:1977pt} defined as
\begin{equation}
S_E(T)= \int d\tau d^3x \left[ \frac{1}{2} \left( \frac{d\Phi}{d\tau} \right)^2 + \frac{1}{2}(\nabla \Phi)^2 + V_{\mathrm{eff}}(\Phi,T) \right]. \nonumber
\end{equation}
Then, $\Gamma = \Gamma_0 {\mathrm e}^{-S_3 / T}$~\cite{Linde:1981zj} and
\begin{align}
S_3(T)
&= \int d^3x \left[ \frac{1}{2} (\nabla \Phi)^2 + V_{\mathrm{eff}}(\Phi,T) \right].
\label{eq_bounceS3}
\end{align}
From the above equations,
in order to obtain the nucleation rate,
the profile of the scalar field $\Phi$ needs to be calculated by solving the following bounce equation:
\begin{align}
\frac{d^2 \Phi}{dr^2} + \frac{2}{r} \frac{d\Phi}{dr} - \frac{\partial V_{\mathrm{eff}}(\Phi,T)}{\partial \Phi}
&= 0,
\end{align}
with the boundary conditions $\frac{d\Phi}{dr}(r=0)=0$ and $\Phi(r=\infty)
=\Phi_{\rm false}$.
The bounce equation can be solved numerically using the overshoot/undershoot method.
The FOPT terminates when nucleation probability of one bubble per horizon volume is of $\mathcal{O}(1)$, i.e., $\Gamma(T_{\ast}) \simeq H_{\ast}^4$.
That is to say, it should satisfy
\begin{equation}\label{tn}
S_3(T_{\ast})/T_{\ast} =4\ln (T_{\ast}/100 \mbox{GeV})+137.
\end{equation}

It is known that there exist three sources for producing GWs during the FOPT, which are collisions of the vacuum bubbles~\cite{Kamionkowski:1993fg}, sound waves~\cite{Hindmarsh:2013xza} and turbulence~\cite{Kosowsky:2001xp, Caprini:2009yp} in the plasma after collisions, respectively.

The most well-known source is the bubbles collisions, and the corresponding phase transition GW spectrum depends on four parameters.
The first parameter is the ratio $\alpha$ of the vacuum energy density released in the phase transition to that of thermal bath, defined as
\begin{equation}
\alpha \equiv \frac{\Delta V_{\mathrm{eff}}(T_{\ast})- T \frac{\partial \Delta V_{\mathrm{eff}}(T_{\ast})}{\partial T}}{\rho_{\rm rad}(T_{\ast})},
\end{equation}
where $\ast$ specifies that the quantity is evaluated at $T_{\ast}$
determined by Eq.(\ref{tn}).
The parameter $\alpha$  measures the strength of the phase transition GWs, namely, larger values for $\alpha$ correspond to stronger phase transition GWs.
The second one is the time duration of the phase transition $\beta^{-1}$ with
$\beta \equiv -\frac{d S_E}{d t} |_{t=t_{\ast}} \simeq\frac{1}{\Gamma}\frac{d \Gamma}{d t} |_{t=t_{\ast}}$, and
one has 
\begin{align}
\frac{\beta}{H_*}
&= \left. T \frac{d(S_3 / T)}{dT} \right|_{T=T_*}
\end{align}
since $\beta = \dot{\Gamma}/\Gamma$ during the  phase transition from its definition.
In other words, $\beta^{-1}$ corresponds to the typical time scale of the phase transition.
The third one is the efficiency factor $\lambda_{co}$, which characterizes the fraction of the energy density  converted into the motion of the colliding bubble walls. And the last one is
the bubble wall velocity $v_{b}$.   
Here, for simplification, we choose the default value $v_b=0.7$.
Explicit calculations on the bubble wall velocity are beyond the scope of this work.
The energy released into the GWs of peak frequency~\cite{Grojean:2006bp} is $\frac{\rho_{GW,co}}{\rho_{tot}}\sim \theta_{co}\left(\frac{H_{\ast}}{\beta}\right)^2\lambda_{co}^2\frac{\alpha^2}{(1+\alpha)^2}v_{b}^{3}.$

The second and third sources are the GWs from the matter fluid effects, which can further contribute to the total energy released in gravitational radiation during the phase transition. Here, we just use the formulae given in Ref~\cite{Caprini:2015zlo}.
The second source is from the sound waves in the fluid, where a certain fraction $\lambda_{sw}$ of the bubble wall energy (after the collision) is converted into motion of the fluid (and is only later dissipated)~\cite{Caprini:2015zlo}
with $\frac{\rho_{GW, sw}}{\rho_{tot}}\sim \theta_{sw}\left(\frac{H_{\ast}}{\beta}\right)\lambda_{sw}^2\left(\frac{\alpha^2}{(1+\alpha)^2}\right).$
The third source is from turbulence in the fluid, where a certain fraction $\lambda_{tu}$ of the walls energy is converted into turbulence~\cite{Caprini:2015zlo} with
$\frac{\rho_{GW, tu}}{\rho_{tot}}\sim \theta_{tu}\left(\frac{H_{\ast}}{\beta}\right)\lambda_{tu}^{3/2}\left(\frac{\alpha^{3/2}}{(1+\alpha)^{3/2}}\right).$
It is worth noticing that these two contributions  from the matter fluid effects depend on $H_{\ast}/\beta$ linearly,
and they are not fully understood.
In some cases, these two effects may be larger than the one from bubble collisions.

The peak frequency produced from bubble collisions at $T_{\ast}$ during the FOPT is given by~\cite{Huber:2008hg, Jinno:2016vai}:
$f_{\rm co}^\ast=0.62\beta/(1.8-0.1v_{b} +v_{b}^{2})$.
Considering the adiabatic expansion of our universe from the early universe to the present universe, the ratio of scale factors at
the time of FOPT and today can be written as $\frac{a_\ast}{a_0}= 1.65 \times 10^{-5} \mbox{Hz}\times\frac{1}{H_{\ast}} \Big( \frac{T_{\ast}}{100 \mbox{GeV}} \Big) \Big( \frac{g^t_\ast}{100} \Big)^{1/6}$.
Thus, the peak frequency today is $f_{\rm co}= f_{\rm co}^\ast a_\ast/a_0$, and the corresponding GW intensity is given by~\cite{Huber:2008hg}
\begin{align}
 \Omega_{\rm co} (f) h^2 \simeq
 &1.67\times 10^{-5} \Big( \frac{H_{\ast}}{\beta} \Big)^2 \Big( \frac{\lambda_{co} \alpha}{1+\alpha} \Big)^2 \Big( \frac{100}{g^t_\ast} \Big)^{\frac{1}{3}} \nonumber \\
 &\times \Big( \frac{0.11v_b^3}{0.42+v_b^3} \Big) \Big[ \frac{3.8(f/f_{\rm co})^{2.8}}{1+2.8(f/f_{\rm co})^{3.8}} \Big]. \nonumber
\end{align}
The peak frequency of the GW signals from sound wave effects is about $f_{\rm sw}=2\beta/({\sqrt{3}v_b}) a_\ast/a_0$ \cite{Hindmarsh:2013xza, Caprini:2015zlo} 
with the GW intensity~\cite{Hindmarsh:2013xza, Caprini:2015zlo,Espinosa:2010hh}
\begin{align}
\Omega_{\rm sw} (f) h^2\simeq \nonumber
& 2.65\times 10^{-6}\Big(\frac{H_{\ast}}{\beta}\Big) \Big(\frac{\lambda_{sw} \alpha}{1+\alpha}\Big)^2
\Big(\frac{100}{g^t_\ast}\Big)^{\frac{1}{3}}v_b\\
&\times\Big[\frac{7(f/f_{\rm sw})^{6/7}}{4+3(f/f_{\rm sw})^2}\Big]^{7/2}, \nonumber
\end{align}
in which
$\lambda_{sw}\simeq \alpha \left(0.73+0.083\sqrt{\alpha}+\alpha\right)^{-1}$~\cite{Espinosa:2010hh} for relativistic bubbles.

The GW signals from the turbulence have the peak frequency at about $f_{\rm tu}= 1.75\beta/v_b~ a_\ast/a_0$\cite{Caprini:2015zlo} and the intensity~\cite{Caprini:2009yp, Binetruy:2012ze}:
\begin{align}
\Omega_{\rm tu} (f) h^2\simeq \nonumber
& 3.35\times 10^{-4}\Big(\frac{H_{\ast}}{\beta}\Big) \Big(\frac{\lambda_{\rm tu} \alpha}{1+\alpha}\Big)^{3/2}
\Big(\frac{100}{g^t_\ast}\Big)^{\frac{1}{3}}v_b\\
&\times\frac{(f/f_{\rm tu})^3}{(1+f/f_{\rm tu})^{11/3}(1+8\pi fa_0/(a_\ast H_\ast))}. \nonumber
\end{align}
The final phase transition spectra consist of the three contributions above.

\section{phase transition gravitational waves from non-Abelian gauge group extended models}
In this section, we discuss the phase transition GWs in some NP models with  extended non-Abelian gauge group, where
one or several strong FOPTs may occur with the evolution of our universe at certain critical temperature.
Firstly, the GW spectra in the gauge group extended models based on the $SU(3)_c \otimes
SU(3)_L \otimes U(1)_Y$ gauge symmetry, commonly known as the 3-3-1
models~\cite{Pisano:1991ee,Frampton:1992wt} are investigated.
The 3-3-1 models can naturally explain the electric charge quantization and three generations of fermions~\cite{Pisano:1991ee,Frampton:1992wt}.
The collider phenomenology of the 3-3-1 models have been extensively studied, such as the recent Ref.~\cite{Cao:2016uur} and references therein,
and the phase transitions in some versions of 3-3-1 models have been studied in Refs.~\cite{Phong:2013cfa,Phong:2014ofa,Borges:2016nne}.
So far, no obvious NP signals are discovered at the LHC, including the 3-3-1 models.
Here, we use the GW signals to explore the NP models and their phase transition patterns
in three versions of the 3-3-1 models (We discuss the minimal and the economical 3-3-1 model in details, and only show
the main results of the reduced
minimal 3-3-1 models.)~\cite{Phong:2013cfa,Phong:2014ofa,Borges:2016nne},
where the scalars fields are accommodated in
different representations of the $SU(3)_L$ gauge group in each version.

For simplicity, we limit our discussions of the FOPT to the thermal barrier case, where the potential barrier in the finite temperature
effective potential origins from thermal effects. In this case, the bosonic fields contribute to the thermal effective
potential of the form $V_{\rm eff} \ni (-T/12 \pi) \bigl( m_{\rm
boson}^2(X,T) \bigr)^{3/2}$ in the limit of high-temperature expansion.
For qualitative sketch of this type of FOPT, we show the general effective potential
near the phase transition temperature, which can be approximated by
\begin{align}\label{vappro}
	V_{\rm eff}(X,T) \sim \frac{\left( - \mu^2 + c \, T^2 \right) X^2 }{2} - \frac{e \, T (X^2)^{3/2}}{12 \pi}  + \frac{\lambda}{4} X^4.
\end{align}
Here, $X$ represents the order parameter field for the phase transition.
For the EW phase transition in the SM, X field is just the Higgs field.
The parameter $e$ quantify the interactions between $X$ field and the light bosons, and
can be schematically written as $e \sim \sum_{\rm light~boson} (\rm degrees~of~freedom) \times (\rm coupling~to~X)^{3/2}$.
And, the parameters $c$ depends on interaction between $X$ and light particles.
For the heavy fields whose masses are much larger than the critical temperature,
their contribution can be omitted from Boltzmann suppression.
This can help to simplify our discussions when the models have many
new fields at different energy scales.
Thus, in this case of qualitative analysis,
the wash out parameter can be obtained as $\frac{\langle X \rangle (T_c)}{T_c} \approx \frac{e}{6 \pi \lambda},$
where the angle bracket $<>$ means the vacuum expectation value (VEV) of the field $X$ at $T_c$.
From the above qualitative analysis, we know that introducing new light bosonic fields (compared to the corresponding critical temperature) helps to produce or enhance the FOPT. The 3-3-1 models just introduce enough bosonic fields to
produce detectable phase transition GWs.

\subsection{Gravitational wave spectrum in the minimal 3-3-1 model}
We firstly consider the phase transition GW spectrum in the so-called minimal 3-3-1 model~\cite{Borges:2016nne}, which
corresponds to the electric charge operator $Q=T_3-\sqrt{3}T_8+x \rm I$.
Here, $x$ represents the $U(1)$ charge, and $T_8$ and $T_3$ are the generators.
The gauge bosons, associated with the gauge symmetry $SU(3)_L\otimes U(1)_Y$,
consist of an octet $W^i_\mu$ ($i = 1,\cdots, 8$) and a singlet $B_\mu$.
In this model, three $SU(3)_L$ triplets scalars ($\eta =\la \eta^{0} \ \eta_{1}^{-} \ \eta_{2}^{+} \rf^T$, $\rho =\la \rho^{+} \ \rho^{0} \ \rho^{++} \rf^T$, $\chi =\la
\chi^{-} \ \chi^{--} \ \chi^{0}\rf^T$) are needed to break the  gauge symmetry,
and generate the masses of the gauge bosons and the exotic quarks.
The scalar potential in terms of $\rho$, $\eta$ and $\chi$ is given as~\cite{Tonasse:1996cx,Borges:2016nne}
\begin{eqnarray}
V\left(\rho,\eta,  \chi\right) &=&
\mu_1^2\eta^\dagger\eta   +
\lambda_1\left(\eta^\dagger\eta\right)^2 +
\mu_2^2\rho^\dagger\rho
\nonumber \\
& + &
\lambda_2\left(\rho^\dagger\rho\right)^2
+ \mu_3^2\chi^\dagger\chi
+
\lambda_3\left(\chi^\dagger\chi\right)^2
\nonumber \\
& + &
\left[\lambda_4\left(\rho^\dagger\rho\right)
  + \lambda_5\left(\chi^\dagger\chi\right)\right]  \left(\eta^\dagger\eta\right)
\nonumber \\
& + &
\lambda_6\left(\rho^\dagger\rho\right)\left(\chi^\dagger\chi\right) +
\lambda_7\left(\rho^\dagger\eta\right)\left(\eta^\dagger\rho\right)
\nonumber  \\ & + &
\lambda_8\left(\chi^\dagger\eta\right)\left(\eta^\dagger\chi\right) +
\lambda_9\left(\rho^\dagger\chi\right)\left(\chi^\dagger\rho\right)
\nonumber \\
& + &
\frac{1}{2} \la f_1\epsilon^{ijk}\eta_i\rho_j\chi_k +
     {\mbox{H. c.}}\rf  \,\,\,.
\label{331m}
\end{eqnarray}
The new gauge bosons acquire masses at several TeV scale when the $SU(3)_L \times U(1)_Y$ group breaks down to
$SU(2)_L \times U(1)_X$ triggered by the $SU(3)_L$ triplet scalar $\chi$,
while the ordinary quarks and SM gauge bosons obtain their masses during
the last step symmetry breaking trigged by the
triplet scalar fields $\eta$ and $\rho$. There exist three CP-even neutral scalars including the lightest one which
corresponds to the SM Higgs boson $h$ and the other heavier scalar bosons
$H_1^0$  and $H_2^0$.  There is also a massive $Z'$ gauge boson, which has been constrained by
the current LHC data.

\begin{figure}
\begin{center}
\includegraphics[scale=0.38]{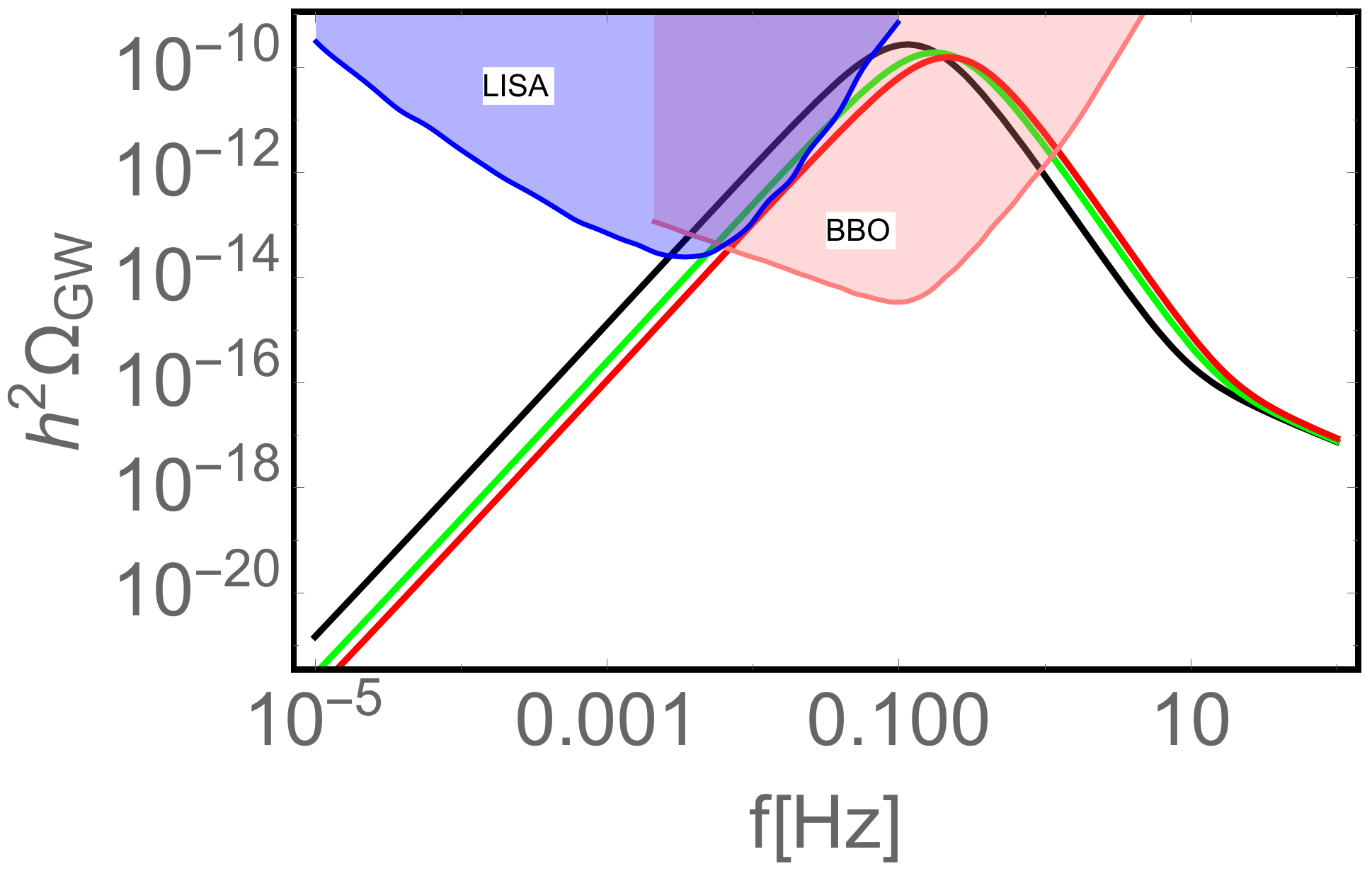}
\caption{The GWs spectra in the minimal 3-3-1 model. The colored regions correspond to the expected sensitivities of GWs interferometers LISA and BBO, respectively. The red line, green line and black line depict the phase transition GW spectrum for the benchmark sets I, II, III in Tab.~\ref{benp}, respectively, during $SU(3)_L \otimes U(1)_Y \to SU(2)_L \otimes U(1)_X$.
}
\label{min}
\end{center}
\end{figure}

Numerically, we find that there are parameter spaces allowed by the collider constraints~\cite{conslhc} that can give a
strong FOPT, when the gauge group spontaneously breaks from
$SU(3)_L \times  U(1)_Y$ to $SU(2)_L\times U(1)_X$~\cite{Borges:2016nne}.
During this phase transition, the order parameter field X here is just the $H_1^0$ field. 
Then, the phase transition GW spectrum can be obtained from the above GW spectrum formulae.
Since this model has so many free parameters,
which makes it very complicated to study the whole
parameter regions allowed, we only show some sets of benchmark points, which are favored by
the collider data and the conditions of a strong FOPT.
Since from the collider constraints (especially the constraints from new gauge boson
$Z'$ at LHC) favor the parameter spaces with $\langle \chi_0 \rangle \gtrsim$ 3~TeV and $Z' > 5$ TeV,
the typical benchmark sets allowed by collider constraints and strong FOPT are shown in Tab.~\ref{benp} with
the corresponding GW signals in Fig.~\ref{min}.
The GW signals include all the contributions from bubble collision, turbulence and sound wave, which are calculated
from Eq.(\ref{veff}) by modifying the package `CosmoTransitions'~\cite{Wainwright:2011kj} .
The detailed discussions on the collider constraints and signals are given in the recent Ref.~\cite{Cao:2016uur} and references therein.
Taking the benchmark set III as an example, when $\langle \chi_0 \rangle =5.1 $ TeV and $m_{Z'} = 5.2$ TeV, the corresponding
GW spectrum is shown as the black line in Fig.~\ref{min}, which is within the sensitivities of LISA and BBO.
The red and green line in Fig.~\ref{min} depicts the GW spectrum for the benchmark set I and II, respectively.
DECIGO and Ultimate-DECIGO can also detect these GW signals.
\begin{table}[h]
\begin{center}
\begin{tabular}{|c|c|c|c|c|c|}
   \hline
   Benchmark set   &$\langle \chi_0 \rangle $[TeV]      &$m_{Z'}$[TeV]    &$(T_{\ast},\alpha, \frac{\beta}{H_*}) $    \\
   \hline
                 I.Red line in Fig.\ref{min}  & 6.1                              & 6.2              & (1.21 TeV, 0,70, 534)                              \\
   \hline
                 II.Green line in Fig.\ref{min}   & 5.4                              & 5.5         & (0.82 TeV, 0.69, 619)                           \\
    \hline
                 III.Black line in Fig.\ref{min}   & 5.1                                & 5.2        & (0.71 TeV, 0.78, 582)                             \\
   \hline
 \end{tabular}
 \end{center}
 \caption{The benchmark sets in the minimal 3-3-1 model for the strong FOPT after considering the constraints from current experimental data.}\label{benp}
\end{table}

It is worth simply discussing what
makes the GW signal from the TeV FOPT available for LISA and significantly larger
than that of the EW phase transition.
It is because the FOPT discussed here comes from the
the potential barrier in the finite temperature
effective potential by thermal effects.
Thus, the phase transition strength is proportional to $e \sim \sum_{\rm light~boson} (\rm degrees~of~freedom) \times (\rm coupling~to~X)^{3/2}$,
namely, the summation of the effective couplings between the order parameter field and the thermal particles (To make efficient thermal contributions, the particle masses should be
much less than $3 T_c$.).
For the EW phase transition, the order parameter field is the Higgs field and only the particles whose masses are smaller
than 1 TeV can make thermal contributions to the EW phase transition. Thus, some new heavy boson in the minimal 3-3-1 model
can not make sufficient contributions to the EW phase transition when their masses are much heavier than the critical temperature
(The critical temperature of EW phase transition is about 100 GeV).
Further, the couplings between the Higgs boson and other particles are greatly constrained by current data, especially the
diphoton decay data, see the detailed discussions on the tensions between strong EW FOPT and LHC data in Ref.~\cite{Chung:2012vg}.
That is why the EW phase transition is rather weak.
For the TeV phase transition, the critical temperature is around 1 TeV, and most of the bosons can make efficient thermal
contributions to the phase transition.
And the order parameter field in TeV phase transition is the new scalar field $H_1^0$, whose collider constraints on the couplings between $H_1^0$ and other particles are not as strong as the Higgs boson case.

\subsection{Phase transition gravitational wave spectra in the economical 3-3-1 model and the reduced minimal 3-3-1 model}

In the economical 3-3-1 model~\cite{Phong:2014ofa}, one chooses the simplest $SU(3)_L$ representations for the scalar fields with spontaneously symmetry breaking,
namely, two complex scalar triplets ($\chi = \left(  \chi^0_1,  \chi^-_2,  \chi^0_3 \right)^T \sim
\left(3,-\frac{1}{3}\right)$ and $\phi = \left(   \phi^+_1,  \phi^0_2,  \phi^+_3 \right)^T\sim
\left(3,\frac{2}{3}\right)$) with different hypercharge are needed.
The scalar potential is written as
\begin{eqnarray}
V(\chi,\phi) &=& \mu_1^2 \chi^\dag \chi + \lambda_1 ( \chi^\dag \chi)^2 + \mu_2^2\phi^\dag \phi  + \lambda_2 ( \phi^\dag\phi)^2  \nonumber \\
 && + \lambda_3 (\chi^\dag \chi)( \phi^\dag \phi) +\lambda_4 (\chi^\dag \phi)( \phi^\dag \chi). \label{veco}
\end{eqnarray}
The $SU(3)_L \otimes U(1)_Y$ gauge group is
broken spontaneously via two steps. In the first step, the symmetry breaking $SU(3)_L \otimes U(1)_Y
\to SU(2)_L \otimes U(1)_X$ happens when
the triplet scalar $\chi$ acquired the VEV given by $\langle\chi\rangle=\frac{1}{\sqrt{2}}\left(  u,  0, \omega\right)^T$
with $\omega \gg v \gg u$.
In the last step, to break into the SM
$U(1)_{\rm EM}$ gauge group $SU(2)_L \otimes U(1)_X \to U(1)_{\rm EM}$, another triplet scalar $\phi$
is needed to acquire the VEV as $\langle\phi\rangle =\frac{1}{\sqrt{2}}\left(
  0,  v,  0 \right)^T.$
In this version of 3-3-1 model, there exist two neutral scalars,
one is the SM Higgs boson
$h$, the other is the heavy scalar $H_1$. It also contains singly charged Higgs boson $H_2^{\pm}$.
There are also two new heavy neutral gauge bosons $Z_2$ and $X_0$, and the
singly charged gauge boson $Y^{\pm}$.
In this work, the modified package `CosmoTransitions'~\cite{Wainwright:2011kj} is used to numerically calculate the FOPT
using full one-loop thermal potential.
During the first time symmetry breaking, the order parameter field for the phase
transition is the $H_1$ scalar field, namely, the $X=H_1$. 
Strong FOPT at the TeV scale  can be induced by the new bosons and exotic quarks
if the masses of these new particle are from $100$ GeV to several TeV.
During the last time symmetry breaking $SU(2)_L \otimes U(1)_X \to U(1)_{\rm EM}$,
the order parameter field for the phase
transition is just the Higgs boson field, namely, $X=h$ if it is compared to Eq.(\ref{vappro}). 
FOPT at the EW scale can be triggered
by the new bosons.
Considering the current constraints from collider data, we have $u < 2.5$ GeV and
 $1~\rm TeV <\omega<5$ TeV~\cite{Phong:2014ofa}. We take conservative
estimation of the parameter spaces for $\omega$ larger than 3 TeV here,
with some reduced parameter spaces shown in Tab.~\ref{bene331}.
The corresponding $(T_{\ast},\alpha, \frac{\beta}{H_*})$
is shown in Tab.~\ref{parae331}.
There is a different aspect from the view point of GW signal
that in the economical 3-3-1 model both the two step symmetry breaking can be the strong
FOPT, which will produce two copies of GWs.

\begin{table}[h]
\begin{center}
\begin{tabular}{|c|c|c|c|c|c|}
   \hline
   Benchmark set   &$\omega $[TeV]      &$m_{H_1}$[TeV]    &$T_{\ast}$[TeV]&    $m_{H_2^{\pm}}$TeV]   \\
   \hline
                 I. Black lines in Fig.3  & 3.0                 & 0.80                 &0.94, 0.16         &        1.4                   \\

    \hline
                 II. Green lines in Fig.3& 4.0                 & 1.30                  &1.26, 0.11         &       2.5                   \\
   \hline
 \end{tabular}
 \end{center}
 \caption{The benchmark sets in the economical 3-3-1 model for two FOPTs after considering the constraints of current experimental data. The $T_{\ast}$ represents the
 corresponding nucleation temperature for the first step FOPT and the second step FOPT, respectively.}\label{bene331}
\end{table}

\begin{table}[h]
\begin{center}
\begin{tabular}{|c|c|c|c|c|c|}
   \hline
   Benchmark set   &$(T_{\ast},\alpha, \frac{\beta}{H_*})$&$(T_{\ast},\alpha, \frac{\beta}{H_*}) $ \\
   \hline
                 I.Black  Fig.4  & (0.94 TeV, 0.59, 305)                                & (0.16 TeV, 0.14, 612)                       \\
      \hline
                 II.Green  Fig.4& (1.26 TeV, 0.68, 413)                                & (0.11 TeV,  0.19,  710)                            \\
   \hline
 \end{tabular}
 \end{center}
 \caption{The corresponding nucleation temperature  $T_{\ast}$, $\alpha$ and $\frac{\beta}{H_*}$  of each FOPT for the different benchmark set in the economical  3-3-1 model.
}\label{parae331}
\end{table}

\begin{figure}
\begin{center}
\includegraphics[scale=0.22]{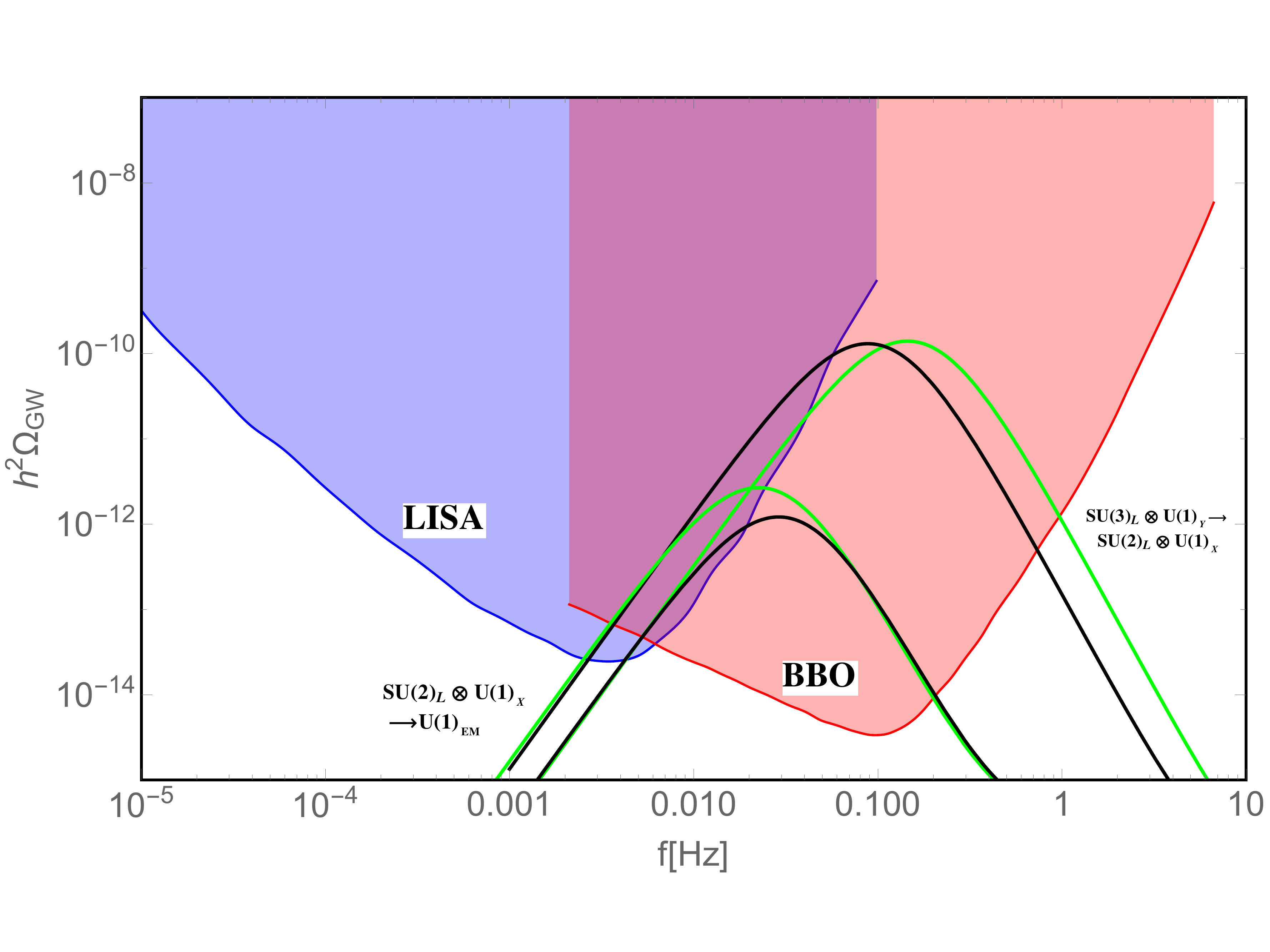}
\caption{The phase transition GW spectra $h^2\Omega_{\rm GW}$ for the benchmark sets in the economical 3-3-1 model. The colored regions correspond to the expected sensitivities of GW interferometers LISA and BBO, respectively. The black lines depict the GW spectra of the benchmark set I for the two FOPTs during $SU(3)_L \otimes U(1)_Y \Longrightarrow SU(2)_L \otimes U(1)_X$ at the TeV (right line) scale and $SU(2)_L \otimes U(1)_X  \Longrightarrow U(1)_{\rm EM}$ at the EW scale (left line), respectively. The green lines  represent the corresponding
GW spectra for the  benchmark set II.
}
\label{eco}
\end{center}
\end{figure}

For the set II of benchmark points with $\omega=4~ \rm TeV$,  $m_{H_1}=1.3$ TeV, $m_{H_2^{\pm}}=2.5$ TeV allowed by current experiments, using the methods and formulae above, these two FOPTs will produce two copies GW spectra with different characteristic peak frequency and amplitude,
as shown by the green lines  in Fig.~\ref{eco}.  The right green line represents the GW signal produced at the first FOPT at TeV scale,
and the left green line depicts the GW signal produced at the second FOPT at EW scale.
It shows that the phase transition GWs can be used to explore this NP model and its phase transition patterns by LISA and BBO. The GW signals are also visible for DECIGO and Ultimate-DECIGO. 
There are also two reasons why the TeV FOPT is stronger than the EW FOPT.
One reason is that the couplings of Higgs boson to other particles in EW phase transition
are greatly constrained by LHC data, and constraints on the couplings of the new scalar bosons to other particles in
TeV phase transition are weaker than the Higgs boson case.
The other reason is that there are more effective thermal particles in the TeV  phase transition case compared to the EW phase transition case since
the critical temperature in the TeV phase transition is obviously higher than the temperature in the EW phase transition.

\begin{figure}
\begin{center}
\includegraphics[scale=0.22]{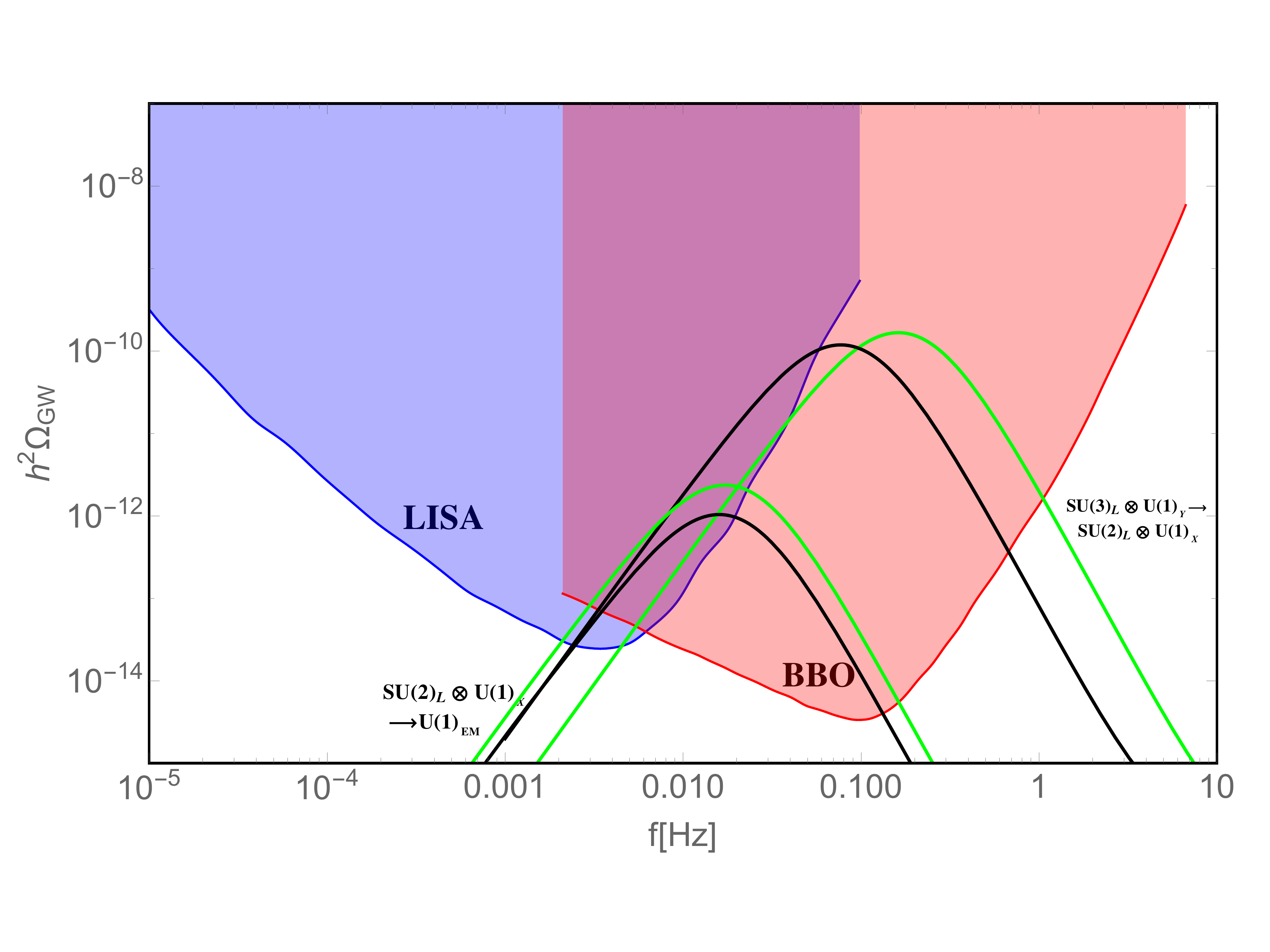}
\caption{The phase transition GW spectra $h^2\Omega_{\rm GW}$ for the benchmark sets in the reduced  minimal 3-3-1 model. The colored regions correspond to the expected sensitivities of GW interferometers LISA and BBO, respectively. The black lines depict the GW spectra of the benchmark set I  for the two FOPTs during $SU(3)_L \otimes U(1)_Y \Longrightarrow SU(2)_L \otimes U(1)_X$ at the TeV scale (right line) and $SU(2)_L \otimes U(1)_X  \Longrightarrow U(1)_{\rm EM}$ at the EW scale (left line), respectively. The green lines  represent the corresponding
GW spectra for the  benchmark set II.
}
\label{rm}
\end{center}
\end{figure}

The GW spectra in the reduced minimal 3-3-1 model is similar to the one of the economical 3-3-1 model since their symmetry breaking and phase transition patterns are similar to the economical model. The reduced minimal 3-3-1 model is mainly composed by
neutral scalars $h$, $H_1$, doubly charged scalar $h^{++}$,
two SM like bosons $Z_1$, $W^{\pm}$, the new heavy neutral boson $Z_2$, the singly and doubly charged boson $U^{\pm\pm}$ and $V^{\pm}$.
These new particles and exotic quarks can be triggers for the strong FOPT~\cite{Phong:2013cfa}.
We show some benchmark sets for the strong FOPT allowed by the collider constraints in Tab.~\ref{benrm331}.
The corresponding $(T_{\ast},\alpha, \frac{\beta}{H_*})$
is shown in Tab.~\ref{pararm331}.
Two copies GW spectra from two FOPTs are produced at different energy scales,
as shown in Fig.~\ref{rm}, which can be detected by LISA and BBO.
DECIGO and Ultimate-DECIGO can also detect these GW signals.
Since from the perspective of the GW signals, the reduced minimal 3-3-1 model has no obvious differences with the economical 3-3-1 model,
we will not discuss this model in detail.
\begin{table}[h]
\begin{center}
\begin{tabular}{|c|c|c|c|c|c|}
   \hline
   Benchmark set   &$\langle \chi_0 \rangle $[TeV]      &$m_{H_1}$[TeV]   &$T_{\ast}$[TeV]&    $m_{h^{++}}$[TeV]   \\
   \hline
                 I.Black  Fig.4  & 3.0                                 & 1.0                 &0.84,  0.080       &        1.9                  \\
      \hline
                 II.Green  Fig.4& 4.0                                 & 1.3                 &1.23,  0.082        &       3.3                   \\
   \hline
 \end{tabular}
 \end{center}
 \caption{The benchmark sets in the reduced minimal 3-3-1 model for the two strong FOPTs after considering the constraints of current experimental data.
 The $T_{\ast}$ represents the
 corresponding nucleation temperature for the first step FOPT and the second step FOPT, respectively.}\label{benrm331}
\end{table}

\begin{table}[h]
\begin{center}
\begin{tabular}{|c|c|c|c|c|c|}
   \hline
   Benchmark set   &$(T_{\ast},\alpha, \frac{\beta}{H_*})$&$(T_{\ast},\alpha, \frac{\beta}{H_*}) $ \\
   \hline
                 I.Black  Fig.4  & (0.84 TeV,0.51,330)                                & (0.08 TeV,0.12,659)                       \\
      \hline
                 II.Green  Fig.4& (1.23 TeV,0.7,490)                                & (0.082 TeV,0.16,719)                            \\
   \hline
 \end{tabular}
 \end{center}
 \caption{The corresponding nucleation temperature  $T_{\ast}$, $\alpha$ and $\frac{\beta}{H_*}$  of each FOPT for the different benchmark set in the reduced minimal 3-3-1 model.
}\label{pararm331}
\end{table}

\subsection{Discussions on gravitational wave spectra in new physics models with  hidden gauge group}
\begin{figure}[h!]
\begin{center}
\includegraphics[scale=0.25]{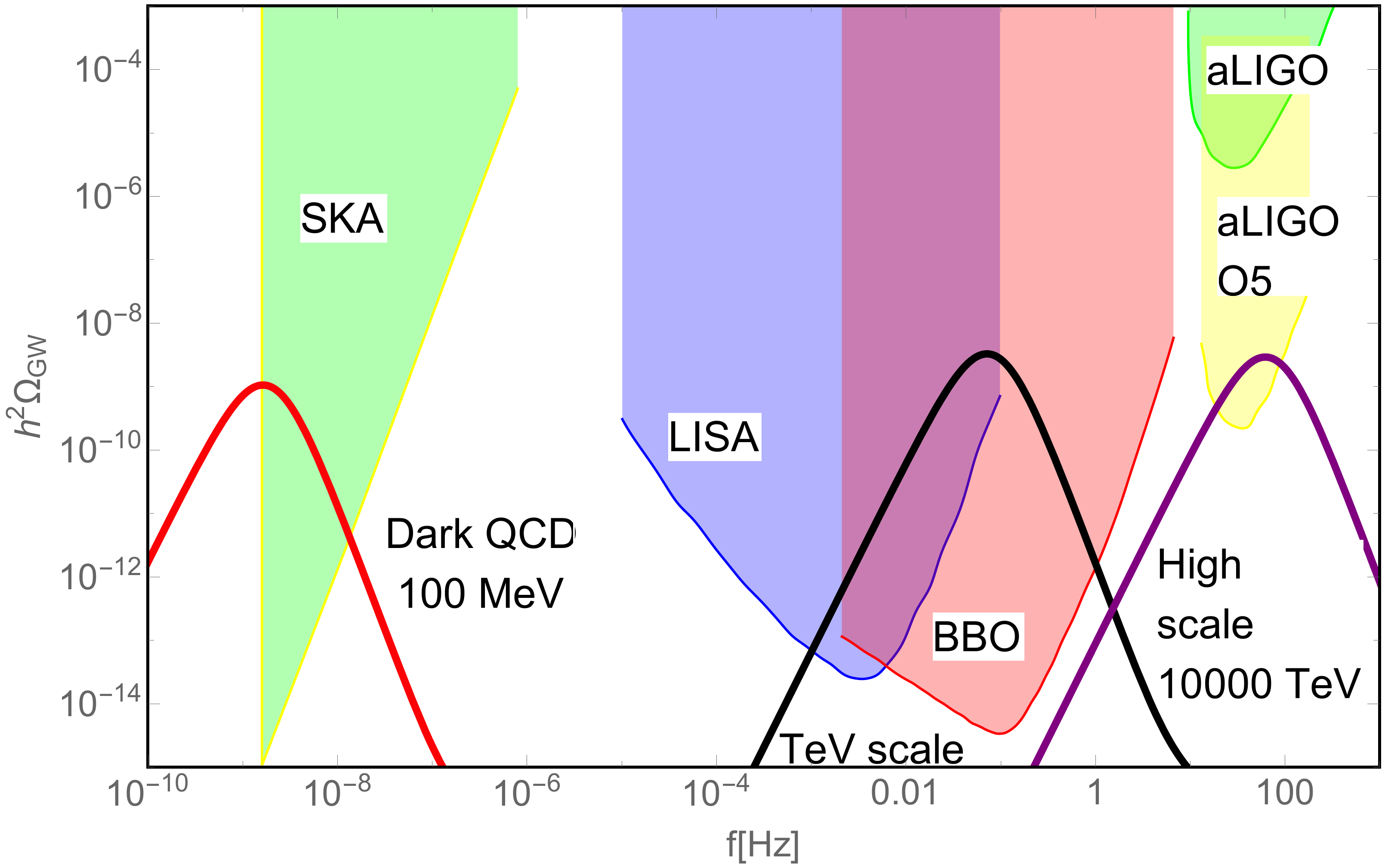}
\caption{Schematic phase transition GW spectra during the evolution of our universe. The colored regions represent the expected sensitivities of GW detectors aLIGO, LISA, BBO and SKA, respectively. The red line depicts the possible GW spectrum if the FOPT occurs at the scale  of $\mathcal{O}(100)$ MeV in some hidden QCD models~\cite{Schwaller:2015tja}. The black line represents the GW spectrum for the FOPT at TeV scale in some  models with extended gauge group. The purple line corresponds to the GW spectrum when the FOPT occurs at the scale of $\mathcal{O}(10000)$ TeV in some NP models with hidden symmetry breaking process.}
\label{gw}
\end{center}
\end{figure}

In general, if the SM is extended by  non-Abelian gauge group, FOPT may occur associated with each step's spontaneously symmetry breaking processes, where the phase transition GWs may be produced and can be used to test the hidden NP models with gauge symmetry breaking.
Thus, the phase transition GWs can be used to test the hidden symmetry breaking during the evolution of the universe.
One class of well-motivated models is the phase transition GW signals in dark matter models with SU(N) hidden gauge group, which are discussed in Ref.~\cite{Schwaller:2015tja}.
If the hidden QCD phase transition scale is about $\mathcal O(100)$ MeV,
the FOPT can produce phase transition GWs with the peak frequency in the $10^{-9}-10^{-7}$ Hz range~\cite{Schwaller:2015tja}, which can be probed by the PTA GW experiments, such as the SKA or FAST.
A schematic GW spectrum for the hidden QCD phase transition is shown in Fig.~\ref{gw} with the red line.
The study in Ref.~\cite{Schwaller:2015tja} may applies to other cases of dark QCD models, such as the case of dark QCD in the famous relaxion mechanism~\cite{Graham:2015cka}\footnote{The relaxion mechanism can technically relax the EW hierarch and the light Higgs mass comes from the dynamical cosmological evolution during the early universe~\cite{Graham:2015cka,Huang:2016dhp}. Especially, the solution to avoid the strong CP problem in the simplest relaxion model, the hidden QCD-like gauge group is
needed, where new dark fermions are also included.
We will study the phase transition GW signals of the relaxion model in
future work~\cite{Huang:2017}.}, and another novel mechanism called ``$N$naturalness"~\cite{Arkani-Hamed:2016rle}\footnote{ The ``$N$naturalness"~\cite{Arkani-Hamed:2016rle} has been proposed to solve the electroweak hierarchy problem
by introducing $N$ copies of the SM with varying values of the Higgs boson mass parameter in a unique universe.
Although, the special reheaton particle has been added to the ``$N$naturalness" mechanism to suppress the baryogenesis from other copies with $v\neq 246$ GeV, the FOPT may still happen in some parameter spaces and produce detectable GWs signals, which will be carefully discussed in our
future work~\cite{Huang:2017}.}.

On the other hand, if a FOPT takes place at a critical temperature of
${\cal O}$($10^7$--$10^8$) GeV~\cite{Dev:2016feu,Balazs:2016tbi}, such as some versions of grand unified models, this could potentially produce detectable GWs spectrum in the future aLIGO or aLIGO-like experiments, and provide us with a unique probe of the hidden NP models at very high energy scale, which is not directly accessible by particle colliders. The schematic GW signal is shown as the purple line in Fig.~\ref{gw}.

\section{Conclusions}
In Fig.~\ref{gw}, the schematic FOPT GW spectra $h^2 \Omega_{\rm GW}$ during the evolution of our universe are shown
for a generic classes of NP models with non-Abelian  symmetry breaking at different energy scales.
The colored regions represent the expected sensitivities of the  GW detectors SKA, BBO, LISA and aLIGO, respectively.
The red line depicts the possible GW spectrum in a class of hidden QCD models~\cite{Schwaller:2015tja},
where the FOPT occurs at the scale of $\mathcal{O}(100)$ MeV and the
associated GWs can be detected by the PTA experiments, such as SKA or the FAST built in China.
The black line represents the GW spectrum for the FOPT at TeV scale in a large classes of NP models with gauge symmetry breaking. As examples, we have shown that three versions of the 3-3-1 models discussed above could produce detectable GWs at TeV scale when the  gauge symmetry $SU(3)_L \otimes U(1)_Y$ breaks to $SU(2)_L \otimes U(1)_X$.
Especially, in the economical and reduced minimal 3-3-1 models, two
FOPTs can take place, which will produce two
copies GW spectra with different characteristic peak frequencies.
In general, large classes of NP models with FOPT at the scale from $\mathcal{O}(100)$ GeV to several TeV can be tested at future laser interferometer GW detectors in space, such as the recently proved LISA~\cite{Audley:2017drz}, BBO, DECIGO, Ultimate-DECIGO, Taiji and TianQin~\cite{Luo:2015ght}.
The purple line corresponds to the GW spectrum in some hidden NP models where the FOPT takes place at the scale of $\mathcal{O}(10000)$ TeV. The GW signals are within the sensitivity of the future aLIGO and provide us with a unique detection of the hidden gauge symmetry breaking at high energy scales beyond the abilities of LHC.

It is worth noticing that this is the first study on that the universe could produce more than
one copies of phase transition GW signal with both solid calculation in realistic NP models (such as two GW
signatures with different characteristics result from two FOPT in the economical and reduced minimal 3-3-1 models) and generic discussions.
Our study also includes the detailed study on phase transition GWs produced at TeV scale in realistic particle physics models.

To conclude, GW signals become a new and realistic approach to explore the
the symmetry breaking patterns in particle cosmology after the discovery of GWs at aLIGO.
For cosmology, we can only hear the non-trivial cosmological phase transitions using GWs to explore the evolution of the universe.
For particle physics, this GW approach can compensate for the colliders, and provide a novel approach to probe the symmetry breaking or
phase transition patterns.
We are in an exciting expedition towards the revolutionary discovery of the NP models and cosmological phase transitions
at GW detectors. More detailed study will be discussed in our future work~\cite{Huang:2017}.

\textit{Acknowledgements.}
We thank  Andrew J. Long, Lian-Tao Wang and Nima Arkani-Hamed for helpful discussions and comments during the workshop at IHEP.
FPH and XZ are supported in part by the NSFC (Grant Nos. 11121092, 11033005, 11375202) and by the CAS pilotB program.
FPH is also supported by the China Postdoctoral Science Foundation under Grant Nos. 2016M590133 and 2017T100108.


\begin{thebibliography}{999}



\bibitem{Abbott:2016blz}
  B.~P.~Abbott {\it et al.} [LIGO Scientific and Virgo Collaborations],
  Phys.\ Rev.\ Lett.\  {\bf 116}, no. 6, 061102 (2016)


\bibitem{Schwaller:2015tja}
  P.~Schwaller,
  Phys.\ Rev.\ Lett.\  {\bf 115}, 181101 (2015)
\bibitem{Dorsch:2014qpa}
  G.~C.~Dorsch, S.~J.~Huber and J.~M.~No,
  Phys.\ Rev.\ Lett.\  {\bf 113}, 121801 (2014)
  doi:10.1103/PhysRevLett.113.121801
  [arXiv:1403.5583 [hep-ph]].
\bibitem{Huang:2016odd}
  F.~P.~Huang, Y.~Wan, D.~G.~Wang, Y.~F.~Cai and X.~Zhang,
  Phys.\ Rev.\ D {\bf 94}, no. 4, 041702 (2016)

\bibitem{Dev:2016feu}
  P.~S.~B.~Dev and A.~Mazumdar,
  Phys.\ Rev.\ D {\bf 93}, no. 10, 104001 (2016)
  doi:10.1103/PhysRevD.93.104001
  [arXiv:1602.04203 [hep-ph]].

\bibitem{Jaeckel:2016jlh}
  J.~Jaeckel, V.~V.~Khoze and M.~Spannowsky,
  Phys.\ Rev.\ D {\bf 94}, no. 10, 103519 (2016)
  doi:10.1103/PhysRevD.94.103519
  [arXiv:1602.03901 [hep-ph]].

\bibitem{Yu:2016tar}
  H.~Yu, B.~M.~Gu, F.~P.~Huang, Y.~Q.~Wang, X.~H.~Meng and Y.~X.~Liu,
  JCAP {\bf 1702}, no. 02, 039 (2017)
  doi:10.1088/1475-7516/2017/02/039
  [arXiv:1607.03388 [gr-qc]].

\bibitem{Addazi:2016fbj}
  A.~Addazi,
  Mod.\ Phys.\ Lett.\ A {\bf 32}, no. 08, 1750049 (2017)
  doi:10.1142/S0217732317500493
  [arXiv:1607.08057 [hep-ph]].

\bibitem{Huang:2016cjm}
  P.~Huang, A.~J.~Long and L.~T.~Wang,
  Phys.\ Rev.\ D {\bf 94}, no. 7, 075008 (2016)
  doi:10.1103/PhysRevD.94.075008
  [arXiv:1608.06619 [hep-ph]].

\bibitem{Huang:2015bta}
  F.~P.~Huang and C.~S.~Li,
  Phys.\ Rev.\ D {\bf 92}, no. 7, 075014 (2015)
  doi:10.1103/PhysRevD.92.075014
  [arXiv:1507.08168 [hep-ph]].

\bibitem{Huang:2015izx}
  F.~P.~Huang, P.~H.~Gu, P.~F.~Yin, Z.~H.~Yu and X.~Zhang,
  Phys.\ Rev.\ D {\bf 93}, no. 10, 103515 (2016)
  doi:10.1103/PhysRevD.93.103515
  [arXiv:1511.03969].

\bibitem{Vaskonen:2016yiu}
  V.~Vaskonen,
  Phys.\ Rev.\ D {\bf 95}, no. 12, 123515 (2017)
  doi:10.1103/PhysRevD.95.123515
  [arXiv:1611.02073 [hep-ph]].

\bibitem{Witten:1984rs}
  E.~Witten,
  Phys.\ Rev.\ D {\bf 30}, 272 (1984).

\bibitem{Hogan:1984hx}
  C.~J.~Hogan,
  Phys.\ Lett.\ B {\bf 133}, 172 (1983);
%
  C.~J.~Hogan,
  Mon.\ Not.\ Roy.\ Astron.\ Soc.\  {\bf 218}, 629 (1986).

\bibitem{Turner:1990rc}
  M.~S.~Turner and F.~Wilczek,
  Phys.\ Rev.\ Lett.\  {\bf 65}, 3080 (1990).

\bibitem{Kamionkowski:1993fg}
  M.~Kamionkowski, A.~Kosowsky and M.~S.~Turner,
  Phys.\ Rev.\ D {\bf 49}, 2837 (1994)
  [astro-ph/9310044].

\bibitem{Hindmarsh:2013xza}
  M.~Hindmarsh, S.~J.~Huber, K.~Rummukainen and D.~J.~Weir,
  Phys.\ Rev.\ Lett.\  {\bf 112}, 041301 (2014)
%


\bibitem{Kosowsky:2001xp}
  A.~Kosowsky, A.~Mack and T.~Kahniashvili,
  Phys.\ Rev.\ D {\bf 66}, 024030 (2002)
  [astro-ph/0111483].




\bibitem{Caprini:2009yp}
  C.~Caprini, R.~Durrer and G.~Servant,
  JCAP {\bf 0912}, 024 (2009)
  [arXiv:0909.0622 [astro-ph.CO]].

\bibitem{Hindmarsh:2015qta}
  M.~Hindmarsh, S.~J.~Huber, K.~Rummukainen and D.~J.~Weir,
  Phys.\ Rev.\ D {\bf 92}, 123009 (2015)


\bibitem{Pisano:1991ee}
  F.~Pisano and V.~Pleitez,
  Phys.\ Rev.\ D {\bf 46}, 410 (1992)
  doi:10.1103/PhysRevD.46.410
  [hep-ph/9206242].

\bibitem{Frampton:1992wt}
  P.~H.~Frampton,
  Phys.\ Rev.\ Lett.\  {\bf 69}, 2889 (1992).
  doi:10.1103/PhysRevLett.69.2889

\bibitem{Seoane:2013qna}
  P.~A.~Seoane {\it et al.} [eLISA Collaboration],
  arXiv:1305.5720.
\bibitem{Audley:2017drz}
  H.~Audley {\it et al.},
  arXiv:1702.00786 [astro-ph.IM].

\bibitem{Corbin:2005ny}
  V.~Corbin and N.~J.~Cornish,
  Class.\ Quant.\ Grav.\  {\bf 23}, 2435 (2006)
  [gr-qc/0512039].


\bibitem{Seto:2001qf}
  N.~Seto, S.~Kawamura and T.~Nakamura,
  Phys.\ Rev.\ Lett.\  {\bf 87}, 221103 (2001)
  doi:10.1103/PhysRevLett.87.221103
  [astro-ph/0108011].

\bibitem{Moore:2014lga}
  C.~J.~Moore, R.~H.~Cole and C.~P.~L.~Berry,
  Class.\ Quant.\ Grav.\  {\bf 32}, no. 1, 015014 (2015)
  doi:10.1088/0264-9381/32/1/015014
  [arXiv:1408.0740 [gr-qc]].

\bibitem{Kudoh:2005as}
  H.~Kudoh, A.~Taruya, T.~Hiramatsu and Y.~Himemoto,
  Phys.\ Rev.\ D {\bf 73}, 064006 (2006)
  doi:10.1103/PhysRevD.73.064006
  [gr-qc/0511145].



\bibitem{Smits:2008cf}
  R.~Smits, M.~Kramer, B.~Stappers, D.~R.~Lorimer, J.~Cordes and A.~Faulkner,
  Astron.\ Astrophys.\  {\bf 493}, 1161 (2009)

\bibitem{Hobbs:2014tqa}
  G.~Hobbs, S.~Dai, R.~N.~Manchester, R.~M.~Shannon, M.~Kerr, K.~J.~Lee and R.~Xu,
  arXiv:1407.0435 [astro-ph.IM].

\bibitem{Graham:2015cka}
  P.~W.~Graham, D.~E.~Kaplan and S.~Rajendran,
  Phys.\ Rev.\ Lett.\  {\bf 115}, no. 22, 221801 (2015)
  doi:10.1103/PhysRevLett.115.221801
  [arXiv:1504.07551].




\bibitem{Linde:1981zj}
  A.~D.~Linde,
  Nucl.\ Phys.\ B {\bf 216}, 421 (1983)
  Erratum: [Nucl.\ Phys.\ B {\bf 223}, 544 (1983)].
  doi:10.1016/0550-3213(83)90293-6, 10.1016/0550-3213(83)90072-X

\bibitem{Coleman:1977py}
  S.~R.~Coleman,
  Phys.\ Rev.\ D {\bf 15}, 2929 (1977)
  Erratum: [Phys.\ Rev.\ D {\bf 16}, 1248 (1977)].
\bibitem{Callan:1977pt}
  C.~G.~Callan, Jr. and S.~R.~Coleman,
  Phys.\ Rev.\ D {\bf 16}, 1762 (1977).
  doi:10.1103/PhysRevD.16.1762

\bibitem{Grojean:2006bp}
  C.~Grojean and G.~Servant,
  Phys.\ Rev.\ D {\bf 75}, 043507 (2007)
  [hep-ph/0607107];

\bibitem{Caprini:2015zlo}
  C.~Caprini {\it et al.},
  JCAP {\bf 1604}, no. 04, 001 (2016)
  doi:10.1088/1475-7516/2016/04/001
  [arXiv:1512.06239].



\bibitem{Huber:2008hg}
  S.~J.~Huber and T.~Konstandin,
  JCAP {\bf 0809}, 022 (2008)
  [arXiv:0806.1828 [hep-ph]].

\bibitem{Jinno:2016vai}
  R.~Jinno and M.~Takimoto,
  Phys.\ Rev.\ D {\bf 95}, no. 2, 024009 (2017)
  doi:10.1103/PhysRevD.95.024009
  [arXiv:1605.01403 [astro-ph.CO]].

\bibitem{Espinosa:2010hh}
  J.~R.~Espinosa, T.~Konstandin, J.~M.~No and G.~Servant,
  JCAP {\bf 1006}, 028 (2010)
  [arXiv:1004.4187].

\bibitem{Binetruy:2012ze}
  P.~Binetruy, A.~Bohe, C.~Caprini and J.~F.~Dufaux,
  JCAP {\bf 1206}, 027 (2012)
  [arXiv:1201.0983 [gr-qc]].


\bibitem{Cao:2016uur}
  Q.~H.~Cao and D.~M.~Zhang,
  arXiv:1611.09337 [hep-ph].



\bibitem{Borges:2016nne}
  J.~S.~Borges and R.~O.~Ramos,
  Eur.\ Phys.\ J.\ C {\bf 76}, no. 6, 344 (2016)
  doi:10.1140/epjc/s10052-016-4168-8
  [arXiv:1602.08165 [hep-ph]].

\bibitem{Phong:2014ofa}
  V.~Q.~Phong, H.~N.~Long, V.~T.~Van and L.~H.~Minh,
  Eur.\ Phys.\ J.\ C {\bf 75}, no. 7, 342 (2015)
  doi:10.1140/epjc/s10052-015-3550-2
  [arXiv:1409.0750 [hep-ph]].

\bibitem{Phong:2013cfa}
  V.~Q.~Phong, V.~T.~Van and H.~N.~Long,
  Phys.\ Rev.\ D {\bf 88}, 096009 (2013)
  doi:10.1103/PhysRevD.88.096009
\bibitem{Tonasse:1996cx}
  M.~D.~Tonasse,
  Phys.\ Lett.\ B {\bf 381}, 191 (1996)
  doi:10.1016/0370-2693(96)00481-9
  [hep-ph/9605230].

\bibitem{conslhc} C.~Salazar, R.~H.~Benavides, W.~A.~Ponce and E.~Rojas,
  JHEP {\bf 1507},  096 (2015).

\bibitem{Chung:2012vg}
  D.~J.~H.~Chung, A.~J.~Long and L.~T.~Wang,
  Phys.\ Rev.\ D {\bf 87}, no. 2, 023509 (2013)
  doi:10.1103/PhysRevD.87.023509
  [arXiv:1209.1819 [hep-ph]].

\bibitem{Wainwright:2011kj}
  C.~L.~Wainwright,
  Comput.\ Phys.\ Commun.\  {\bf 183}, 2006 (2012)
  doi:10.1016/j.cpc.2012.04.004
  [arXiv:1109.4189].


\bibitem{Arkani-Hamed:2016rle}
  N.~Arkani-Hamed, T.~Cohen, R.~T.~D'Agnolo, A.~Hook, H.~D.~Kim and D.~Pinner,
  Phys.\ Rev.\ Lett.\  {\bf 117}, no. 25, 251801 (2016)
  doi:10.1103/PhysRevLett.117.251801
  [arXiv:1607.06821 [hep-ph]].

\bibitem{Quiros:1999jp}
  M.~Quiros,
  hep-ph/9901312.


\bibitem{Huang:2016dhp}
  F.~P.~Huang, Y.~Cai, H.~Li and X.~Zhang,
  Chin.\ Phys.\ C {\bf 40}, no. 11, 113103 (2016)
  doi:10.1088/1674-1137/40/11/113103
  [arXiv:1605.03120 [hep-ph]].




\bibitem{Huang:2017}
  F.~P.~Huang,  and X.~Zhang,
 work in progess.


\bibitem{Balazs:2016tbi}
  C.~Balazs, A.~Fowlie, A.~Mazumdar and G.~White,
  arXiv:1611.01617 [hep-ph].

\bibitem{Luo:2015ght}
  J.~Luo {\it et al.} [TianQin Collaboration],
  Class.\ Quant.\ Grav.\  {\bf 33}, no. 3, 035010 (2016)
  doi:10.1088/0264-9381/33/3/035010
  [arXiv:1512.02076 [astro-ph.IM]].


















\end{thebibliography}
\end{document}